\documentclass[12pt,a4paper]{article} 
\pdfoutput=1
\usepackage{jheppub}
\usepackage{verbatim}

\usepackage{enumitem} 
\usepackage{amssymb,amsmath,multirow,slashed,graphicx} 

\usepackage{graphicx}
\usepackage{lipsum}
\usepackage{bm}
\usepackage{setspace}
\usepackage{cancel}
\usepackage[normalem]{ulem}

\newcommand{\be}{\begin{equation}}
\newcommand{\ee}{\end{equation}}
\newcommand{\bea}{\begin{eqnarray}}
\newcommand{\eea}{\end{eqnarray}}
\newcommand{\bal}{\begin{aligned}}
\newcommand{\eal}{\end{aligned}}

\def\Eq#1{Eq.~(\ref{#1})}

\begin{document}

\title{pNGB Higgs Naturalness at a Tipping Point}

\author[a]{Matthew McCullough,}
\affiliation[a]{Theoretical Physics Department, CERN, 1211 Geneva 23, Switzerland}

\author[b]{Adriana Menkara}
\affiliation[b]{Deutsches Elektronen-Synchrotron DESY, Notkestr.~85, 22607 Hamburg, Germany}

\author[c]{and Ennio Salvioni}
\affiliation[c]{Department of Physics and Astronomy, University of Sussex, Sussex House,\\ BN1 9RH Brighton, United Kingdom}

\abstract{
In scenarios where the Higgs is viewed as a pseudo Nambu-Goldstone boson (pNGB) the question of naturalness finds itself, from a phenomenological perspective, at a tipping point between direct searches and precision. If, by the end of the High-Luminosity LHC operation, all experimental results were to remain consistent with the Standard Model, precision Higgs coupling measurements will begin to drive the naturalness tension. To illustrate this from a fresh perspective we construct a maximally natural `Kitchen Sink' model, throwing into the mix three approaches to symmetry-based naturalness: Supersymmetry, Twin Higgs, and pNGB Higgs models with a Gegenbauer potential. In other words, we build a `Supersymmetric Gegenbauer's Twin' model. This model not only maximises naturalness, at least from a technical perspective, but can also interpolate between all three ingredients smoothly, revealing the interplay between direct exploration and precision. Implications for FCC-ee and FCC-hh are discussed.
}

\preprint{CERN-TH-2025-067, DESY-25-052}

\maketitle


\section{Introduction} \label{sec:intro}
One of the most pressing challenges in particle physics is to determine the fundamental origins of the Standard Model (SM), which the Higgs sector hints must be ultraviolet (UV) completed at a nearby energy scale. At present four observations that appear, at face value, self-evident also seem contradictory:
\begin{enumerate}[label=(\roman*)]
\item  The SM is not UV complete and so must be replaced with some more fundamental UV description which predicts the SM parameter values.  
\item  This UV completion will comprise new states at a new mass scale.  
\item  At the scale where this new theory is matched to the SM effective field theory the Higgs field will receive proportionate mass contributions, as there is no symmetry present at this scale to say otherwise.  
\item  Searches for coloured particles at the LHC have reached up to and beyond the TeV scale and nothing beyond the SM has been observed.
\end{enumerate}
The apparent inconsistency of these statements is known as the naturalness puzzle, whose get-out clause is the possibility of fine-tuning the UV parameters against one another in order to realise a separation between the UV scale and the electroweak (EW) scale. The price of this fine-tuning is not only its implausibility but, even more importantly, the loss of predictivity that would ultimately result from it, see point (i).

As concerns phenomenology, there are two faces to the question of Higgs naturalness: direct searches for new states and measurements of Higgs couplings. As of now the majority of LHC analyses are based on $140 \text{ fb}^{-1}$ of data, whereas in the future $3 \text{ ab}^{-1}$ are anticipated. Thus we expect the sensitivity to the best-measured Higgs couplings, with small systematic uncertainties, to improve by up to a factor $(3000/140)^{1/2} \approx 4.6$ between now and the end of the HL-LHC era. For direct searches, one finds (by employing \texttt{Collider Reach}~\cite{ColliderReach} to make crude estimates) that for states whose masses have lower bounds in the $0.5\,$-$\,3$ TeV range presently, an improvement in the mass constraint by at most a factor $1.5\,$-$\,1.9$ is expected, assuming production by gluons.\footnote{As they are more model-dependent we do not consider additional ancillary probes of natural models for the Higgs sector, such as flavour and precision EW observables which, depending on assumptions regarding the UV, can be powerful. See for instance~\cite{Glioti:2024hye}.}

In this work we consider one of the classes of UV completions of the SM which have dominated discourse on naturalness in the past, namely pNGB Higgs models (see~\cite{Panico:2015jxa} for a review). In these models the fine-tunings $\Delta$ required within the UV to realise the observed Higgs vacuum expectation value and mass in the infrared (IR) typically scale as
\be
\Delta ~~ \propto ~~ \delta_{hXX} \;, \quad \frac{m_h^2}{M_R^2} \;,
\ee
where $\delta_{hXX} \equiv g_{hXX}/g_{hXX}^{\rm SM} - 1$ characterises the magnitude of modifications to Higgs couplings in the IR and $M_R$ denotes the mass scale of new resonances.  Thus we expect that, were all measurements to remain SM-like until the end of the High-Luminosity LHC (HL-LHC) operation, the impact of precision Higgs coupling measurements on the picture of naturalness for these models will grow as compared to the impact of direct searches for new resonances.  More quantitatively, we expect the relative reach of the two complementary probes to scale, very roughly, as
\bea \label{eq:probes_hierarchy}
\frac{\Delta_{\text{Indirect}}}{\Delta_{\text{Direct}}} \bigg|_{\frac{3 \text{ ab}^{-1}}{140 \text{ fb}^{-1}}}
 \,\approx\,  \frac{1.7^2}{4.6}\,  \approx\,  0.6
\eea
between now and then, if no evidence for beyond the SM physics arises in the meantime. Given the mild hierarchy in Eq.~\eqref{eq:probes_hierarchy}, which class of observables will ultimately dominate the picture of naturalness depends on the details of explicit models.  In this section we attempt to tread the precipitous ridgeline between generality and specificity in order to answer this question informatively. To this end we consider, beyond pNGB Higgs models, also supersymmetric (SUSY) models for the sake of comparison.

We begin by discussing the relation between fine-tuning and direct searches for new resonances. We frame our exposition in the context of pNGB Higgs models, but note that for this aspect the same conclusions apply to SUSY. In general, a pNGB can be naturally light \emph{without} the presence of any other states at a similar mass scale, as a consequence of its approximate shift symmetry; for the Higgs, the shift symmetry must be extended to a non-Abelian realisation to accommodate the EW interactions. Then, regardless of the nature of the UV completion, any SM interactions which cannot respect the shift symmetry generically give rise to contributions to the Higgs potential in the UV. This UV-sensitivity only disappears above the mass scale of new states, which allow for the realisation of extended UV symmetries. 

In this perspective, the most consequential coupling of the Higgs boson is the one to the top quark. In any UV completion of the pNGB Higgs one should expect, on symmetry grounds, contributions to the mass of the form
\be
\delta m_h^2 \approx  \frac{3 y_t^2}{4 \pi^2} M_T^2 \;,
\ee
where $M_T$ is the mass scale at which the interaction becomes symmetric, corresponding to the masses of the fermionic top partners. In SUSY, $M_T$ would stand for the mass scale of the top squarks. In both classes of UV completions, assuming that the symmetry-restoring states are coloured the results of LHC searches to date imply a minimal fine-tuning contribution of around
\be
\Delta_{m_h^2} \lesssim 0.15 \left(\frac{1.2 \text{ TeV}}{M_T} \right)^2 \;.
\ee
In realistic scenarios the prefactor can be smaller and so the fine-tuning can be worse. However, the important point here is not so much the precise absolute value of the fine-tuning, but rather how that fine-tuning will evolve as we enter and ultimately complete the HL-LHC phase, hopefully followed by a Higgs factory era. The fine-tuning arising from the non-observation of new coloured states scales quadratically with their mass. Hence the increase in fine-tuning, under the assumption of continued SM-like observations, between now and the end of the HL-LHC would likely be in the ballpark of $\sim 2.9$. Accordingly, we expect the irreducible mass fine-tuning to reach around the $5\%$ level.  

We now turn to the question of fine-tuning and Higgs coupling measurements, where pNGB and SUSY scenarios show important differences. Consider first the Minimal SUSY SM, where the scalar sector has a two-Higgs-doublet-model (2HDM) structure. An indirect effect of the stops is the modification of the $hgg$ coupling relative to the SM: if stop mixing can be neglected, one finds an enhancement
\bea
\delta_{hgg} \approx   \frac{m_t^2}{2 m^2_{\tilde{t}}} \, \approx\,  1\%\,  \left(\frac{1.2 \text{ TeV}}{m_{\tilde{t}}} \right)^2 \;,
\eea
where $m_{\tilde{t}}$ is the stop mass. However, a comparison to the HL-LHC projection $|\delta_{hgg}| < 4\%$ (at $2\hspace{0.2mm}\sigma$ confidence~\cite{deBlas:2019rxi}) shows that the sensitivity is not competitive with direct stop searches. The fine-tuning related to the 2HDM structure has recently been reassessed in \cite{Bernal:2022wct}. We assume there is a sufficiently large mass hierarchy that we may integrate out the heavy Higgs doublet and work in the resulting decoupling limit.  In \cite{Bernal:2022wct} it was found that the fine-tuning of the vev is, approximately,
\be \label{eq:vev_tuning_2HDM}
\Delta_{v^2} \approx \frac{m_h^2}{4m_A^2} \tan^2 \beta\;,
\ee
where $m_A \gg m_Z$ is the mass of the heavy pseudoscalar boson, reflective of the mass scale of the entire heavy Higgs doublet. In the same limit the modifications of the Higgs couplings to fermions are (see e.g.~\cite{Djouadi:2005gj})
\bea \label{eq:hff_MSSM}
\delta_{h\overline{t}t}  \approx  \frac{2 m_Z^2}{m_A^2} \cos^2 \beta \cos 2 \beta\;,  \qquad \delta_{h\overline{b}b}  \approx  -\,\frac{2 m_Z^2}{m_A^2} \sin^2 \beta \cos 2 \beta\;, 
\eea
whereas the coupling to gauge bosons arises at the next order in the $m_Z^2/m_A^2 \ll 1$ expansion, $\delta_{hVV} \approx -\, m_Z^4 \sin^2 4 \beta / (8 m_A^4)$~\cite{Djouadi:2005gj}. Combining Eqs.~\eqref{eq:vev_tuning_2HDM} and~\eqref{eq:hff_MSSM} yields
\be
\Delta_{v^2} \approx \delta_{h\overline{t}t}\; \frac{\tan^2 \beta}{4 \cos^2 \beta \cos 2 \beta}  \;,\qquad \Delta_{v^2} \approx  -\, \delta_{h\overline{b}b} \;  \frac{1}{4 \cos^2 \beta \cos 2 \beta} \;.
\ee
We thus observe a relation between Higgs coupling deviations and fine-tuning, in the general and well-known sense that the more SM-like the Higgs is, the more fine-tuned it must be. Ultimately, more precise Higgs coupling measurements would lead to an increase in tuning of $\sim 4.6$ if they remain consistent with the SM. However, due to the strong $\tan \beta$ dependence it is difficult to argue about the absolute scale of fine-tuning implied by Higgs coupling measurements. Furthermore, the model-dependence goes beyond $\tan\beta$: for instance, there exist SUSY scenarios in which the 2HDM structure is evaded and the Higgs couplings at tree-level are exactly SM-like~\cite{Davies:2011mp}. In conclusion, we are unable to draw a robust conclusion on the connection between fine-tuning and Higgs couplings in SUSY, and therefore on the relative power of direct versus indirect probes of naturalness in this context.

Next, we consider fine-tuning and Higgs couplings in pNGB Higgs scenarios (see for instance~\cite{Giudice:2007fh,Panico:2015jxa}). In these models some continuous global symmetry group $\mathcal{G}$ is spontaneously broken to a subgroup $\mathcal{H}$, with the pNGBs furnishing the coset $\mathcal{G}/\mathcal{H}$. This coset must be able to accommodate the SM Higgs doublet and $\mathcal{H}$ must contain the EW $\mathrm{SU}(2)_L \times \mathrm{U}(1)_Y$ as a subgroup, so that it can be gauged. Explicit breaking of $\mathcal{G}$ is necessary for a realistic scenario and leads to a vacuum misalignment angle $\sin ( \langle h \rangle/f ) = v/f$, where $v\approx 246$~GeV and $f$ is the scale of spontaneous $\mathcal{G} \to \mathcal{H}$ breaking. As a result, the $hVV$ couplings are suppressed relative to the SM by
\be
\frac{g_{hVV}}{g_{hVV}^{\rm SM}} \,=\, \cos (\langle h \rangle / f) \, \approx \, 1- \frac{v^2}{2 f^2}  \;.
\ee
Thus, in a pNGB scenario we expect modifications of the Higgs couplings whose magnitude is tightly tied to the UV compositeness scale.

If one assumes all of the explicit symmetry breaking arises due to the SM gauge and Yukawa couplings, then the leading contributions to the Higgs potential are of the form
\be
V \approx \beta f^2 \Lambda^2 \left(- \,r \sin^2 \frac{h}{f}+\sin^4 \frac{h}{f} \right) ~~,
\ee
where $\Lambda$ is a composite mass scale, $r$ is a priori expected to be an $\mathcal{O}(1)$ coefficient and $\beta$ is typically around a top-loop factor \cite{Panico:2015jxa}.  The observed Higgs vev is found at
\be
2\sin^2 \frac{\langle h \rangle}{f} = r  ~~.
\ee
There is typically no symmetry in this theory which can suppress the coefficient of one term of $V$ over the other, therefore obtaining a small $v/f$ requires that parameters in the UV theory are fine-tuned so as to largely cancel and give a small value for $r$.  This fine-tuning issue is a generic problem for pNGB Higgs models and is known as `$v/f$'-tuning \cite{Panico:2015jxa}.  Ultimately, in these models
\bea
\Delta_{v^2}  \approx  \frac{2 v^2}{f^2} \,\approx\,  4 | \delta_{hVV} | \;,
\eea
hence the present ATLAS and CMS sensitivity $|\delta_{hVV}| \lesssim 3.5\%$ ($1\hspace{0.2mm}\sigma$,~\cite{ATLAS-CONF-2021-053,CMS-PAS-HIG-19-005}) leads to an irreducible tuning of around $15\%$, which would ultimately evolve to about $5\%$ by the end of the HL-LHC phase were no Higgs coupling modifications observed ($|\delta_{hVV}| < 1.3\%$~\cite{deBlas:2019rxi}). With the precision on Higgs coupling measurements anticipated at FCC-ee, $|\delta_{hVV}| < 0.17\%$~\cite{deBlas:2019rxi}, this tuning would be pushed below the percent level.

Factors of $\mathcal{O}(1)$ matter, hence this discussion can only be considered qualitative. Nonetheless it does suggest that, if LHC results continue to agree with the SM, Higgs coupling measurements may come to lead the naturalness picture for pNGB Higgs models. In a scenario where the HL-LHC is followed by a Higgs factory, Higgs coupling measurements will ultimately dominate the naturalness conversation. We will return to these points in Sec.~\ref{sec:Outlook}.

\section{A `Kitchen Sink' Model}
In attempting to characterise the phenomenological landscape broadly, the discussion in Sec.~\ref{sec:intro} leaves loopholes. In particular, while it appears impossible to disentangle the mass of symmetry-completing heavy new resonances from the question of Higgs naturalness, their phenomenology may be exotic. The coupling-modification connection seems potentially pliable also.  To this end, we attempt to construct a maximally natural UV completion in which the most relevant IR observables are calculable as a function of the underlying microscopic parameters, thus placing naturalness discussions on a sure footing, yet the phenomenological signatures are diluted as much as possible.  From this perspective we will consider the possible evolution of naturalness. The philosophy will be to not inject \ae sthetic model-building elements which may, or may not, also be warranted, but to adhere to technical naturalness alone, such that the theory is radiatively stable, with stabilising symmetries being restored in the limit where small symmetry-breaking parameters vanish.  Our model will be a concoction of three naturalness-motivated ingredients:  Supersymmetry, Twin Higgs, and pNGB Higgs scenarios implementing the Gegenbauer mechanism~\cite{Durieux:2021riy,Durieux:2022sgm}; we refer to it as the `Kitchen Sink' model since everything but the kitchen sink of symmetry-based naturalness scenarios is thrown at the problem.

We will interpret the Higgs boson as a pNGB of a spontaneously broken approximate global symmetry.  The symmetry is approximate due to some small parameters in the underlying theory which do not respect the global symmetry. The power of the effective field theory (EFT) approach is that, for a given pattern of spontaneous and explicit symmetry breaking, the broad universality class of possible microscopic UV completions gives rise to the same IR EFT.

The two irreducible sources of explicit global symmetry breaking for a pNGB Higgs, and therefore of its potential, are the SM Yukawa interactions and EW gauge interactions. The Yukawa coupling of the top quark typically gives the dominant effect. Taking as example the spontaneous symmetry breaking pattern $\text{SO}(5) \to \text{SO}(4)$, known as the Minimal Composite Higgs model~\cite{Agashe:2004rs} as it is the smallest coset that includes custodial symmetry, the top sector generates a potential for the Higgs of the form
\be
V_t \approx \frac{N_c y_t^2}{16 \pi^2} f^2 M_T^2  \left(a_2 \cos \frac{2h}{f} + a_4  \cos \frac{4 h}{f} \right) ~~,
\label{eq:toppot}
\ee
where $M_T$ is the mass scale associated with heavy coloured fermions which UV-complete the top Yukawa interaction and the $a_j$ are expected to be $\mathcal{O}(1)$ coefficients.

\subsubsection*{Electroweak scale}
The $y_t\,$-$\,$generated potential in Eq.~\eqref{eq:toppot} is typically minimised at $\langle h \rangle = 0$, which is phenomenologically unacceptable, or at $\langle h \rangle \sim f$.  The latter is also phenomenologically unacceptable because there is no separation of the compositeness and EW scales, giving rise to $\mathcal{O}(1)$ modifications of Higgs properties. As discussed above, it is usually assumed that a viable EW scale is instead generated by having multiple contributions to the potential of this form, be they of gauge origin or arising from different fermion multiplets, and fine-tuning the coefficients so as to generate a global minimum at $\langle h \rangle \ll f$. 

In \cite{Durieux:2021riy} an alternative possibility was proposed: that in the UV there is an additional source of explicit symmetry breaking, associated to a larger $\text{SO}(5)$ irrep. The possibility of multiple sources of explicit breaking is inspired by the pions of QCD, which are subject to sources of explicit symmetry breaking of uncorrelated origin in the UV, namely the QED gauge interactions and the quark masses. It was shown in \cite{Durieux:2021riy} that for a spurion in an $n$-index traceless symmetric irrep of $\text{SO}(5)$, which realises the explicit breaking $\mathrm{SO}(5) \to \mathrm{SO}(4)$, the pNGB Higgs potential receives the contribution
\be
V_G^{(n)} = \epsilon f^2 M^2 G_n^{3/2} \Big( \hspace{-0.75mm}\cos \frac{h}{f} \Big) ~~,
\ee
where $\epsilon$ is a small dimensionless parameter, $M$ is the UV scale the spurion originates from and $G_n^{3/2}$ is a Gegenbauer polynomial. The possibility of large $n$ (significantly larger than the $n = 2, 4$ values typically found in the top sector potential of Eq.~\eqref{eq:toppot}) is motivated phenomenologically since, due to the shape of the Gegenbauer polynomials, it can naturally generate a global minimum with parametrically small $v/f$, removing this source of UV parameter fine-tuning~\cite{Durieux:2021riy}.

\subsubsection*{Higgs mass}
A further observation concerning \Eq{eq:toppot} is that its typical prediction for the Higgs boson mass is $m_h^2 \approx N_c y_t^2 M_T^2 / (4 \pi^2)$~\cite{Panico:2015jxa}. Since $M_T \lesssim 450$~GeV has long been ruled out by LHC searches for new coloured fermions, an additional fine-tuning beyond the one required to realise $v\ll f$ is necessary, to explain why the Higgs is light. The `Gegenbauer Higgs' proposal of \cite{Durieux:2021riy} does nothing to address this light Higgs boson fine-tuning.

To explain how the Higgs could be naturally light within this context, it was thus proposed in \cite{Durieux:2022sgm} to wed the Gegenbauer scenario with the `Twin Higgs' idea of \cite{Chacko:2005pe,Barbieri:2005ri,Chacko:2005vw,Chacko:2005un}. In basic terms, within the Twin Higgs paradigm the role of the coloured fermions is, for some range of scales, instead played by SM-gauge-neutral fermions. Thus the `Gegenbauer's Twin' proposal of~\cite{Durieux:2022sgm} realises a naturally light Higgs boson, due to the twinning, and a naturally small EW scale, due to the Gegenbauer contribution.

\subsubsection*{Strong versus weak coupling}
As emphasised in \cite{Barbieri:2015lqa,Contino:2017moj} (see also \cite{Geller:2014kta,Craig:2015pha,Low:2015nqa}), in a Twin Higgs scenario the radiative contributions to the Higgs potential from the top sector are modified as compared to an un-twinned scenario, becoming approximately of the form
\be\label{eq:V_SM}
V_t \approx \frac{N_c y_t^4}{64 \pi^2} f^4  \bigg[ \sin^4 \frac{h}{f}\, \log \frac{a}{\sin^2 \frac{h}{f} } + \cos^4 \frac{h}{f}\, \log \frac{a}{\cos^2 \frac{h}{f} } \bigg] \;,
\ee
where $y_t = y_t^{\overline{\mathrm{MS}}}(m_t) \approx 0.94$. The parameter $a$ is UV-dependent and incalculable from within the IR theory.  However, in general one expects that
\be\label{eq:a}
a \approx \frac{ g_\ast^2 }{ y_t^2 } \;,
\ee
where $g_\ast$ is the coupling associated with new coloured resonances in the top sector. In a standard composite Twin scenario one finds that, for the observed EW scale and Higgs mass, one requires $g_\ast \sim 4\pi$, thus fitting comfortably within the context of a strongly coupled UV completion \cite{Barbieri:2015lqa}. This comes, however, at the price of the $v/f$ fine-tuning required. On the other hand, in the `Gegenbauer's Twin' scenario, in parameter regions where $v\ll f$ is naturally generated by the Gegenbauer potential, one instead typically requires $a \lesssim 2$: for example, for a representative parameter point with $f=1$ TeV and Gegenbauer index $n=6$ a value $a \approx 1.3$ is needed~\cite{Durieux:2022sgm}.

Two comments are in order. The first is to urge some caution over a too-strict interpretation of \Eq{eq:a}, since the logs are also related to the scale of threshold effects in the UV completion, hence \Eq{eq:a} should be considered as a guide, rather than a strict relation. That being said, the relatively small value of $a$ required does suggest that while the `Gegenbauer's Twin' scenario offers a natural UV completion of the SM Higgs sector, it likely finds its own UV embedding within perturbative, rather than strongly coupled, completions.

We now pursue a compelling candidate for a perturbative UV completion of the Twin Higgs scenario: a SUSY Twin Higgs~\cite{Falkowski:2006qq,Chang:2006ra,Craig:2013fga,Katz:2016wtw,Badziak:2017syq,Badziak:2017kjk,Badziak:2017wxn}. Compelling because such theories are valid down to very small distance scales, are conducive to perturbative analysis, and will automatically realise $a\sim 1$ since the top-sector resonances interact with strength fixed by the top Yukawa coupling. We will discuss to what extent adding a Gegenbauer contribution to the SUSY Twin Higgs, thus realising a `Supersymmetric Gegenbauer's Twin' scenario, can maximise naturalness by alleviating the $v/f$ tuning.

\subsection*{The model}
In the SUSY Twin Higgs~\cite{Falkowski:2006qq,Chang:2006ra,Craig:2013fga,Katz:2016wtw,Badziak:2017syq,Badziak:2017kjk,Badziak:2017wxn} one has two copies of the MSSM, with equality of masses and couplings enforced by an approximate exchange symmetry under which the fields of both sectors are swapped.  We refer to the `visible' MSSM as $A$ sector and the Twin MSSM as $B$ sector. In addition, we introduce a singlet superfield $\mathcal{S}$ that couples the two sectors.

The $\mathcal{Z}_2$ and gauge symmetries ensure that both the superpotential and the soft masses accidentally preserve an SO(8) global symmetry. We make this explicit by pairing the Higgs doublets of the two sectors, $h^{A}_{u,d}$ and $h^{B}_{u,d}$ charged under $[\text{SU}(2) \times$U(1)]$_{A,B} \subset \text{SO}(4)_{A,B}$, into multiplets transforming in the fundamental of $\text{SO}(8)$ as $\phi_{u,d} = (h^{A}_{u,d}, h^{B}_{u,d})$. In this notation the superpotential reads
\be
    W = \frac{ \mu }{\sqrt{2}} \phi_u \phi_d + \frac{ \lambda}{2} \mathcal{S} \phi_u \phi_d + \frac{M_{\mathcal{S}}}{2}  \mathcal{S}^2 ~,
\ee
while the soft masses are
\begin{equation}
V_{\rm soft} = \frac{ m^2_{H_u}}{2} |\phi_u|^2 + \frac{m^2_{H_d}}{2} |\phi_d|^2 - \frac{b}{2} \left( \phi_u \phi_d + \mathrm{h.c.}\right) + m_{\mathcal{S}}^2 |\mathcal{S}|^2\,.
\end{equation}
Assuming $m_{\mathcal{S}}^2 \gg M_{\mathcal{S}}^2, \mu^2$, integrating out $\mathcal{S}$ leaves a non-decoupling $F$-term quartic,
\begin{equation}
    V_{\text{SO}(8)} = \frac{1}{2} \left(\mu^2 + m_{H_u}^2 \right) \lvert \phi_u \rvert^2 + \frac{1}{2} \left(\mu^2 + m_{H_d}^2 \right) \lvert \phi_d \rvert^2 - \frac{b}{2}\left( \phi_u \phi_d + \mathrm{h.c.}\right) + \frac{\lambda^2}{4} \lvert  \phi_u \phi_d\rvert ^2  ~.
\end{equation}
Minimisation of this potential determines $\tan \beta \equiv v_u^A/v_d^A = v_u^B / v_d^B$ and the total magnitude of the SO(8)-breaking vev $f^2 \equiv (v_u^A)^2 + (v_d^A)^2 + (v_u^B)^2 + (v_d^B)^2$, where $v_{u,d}^{A,B} = \sqrt{2}\, \langle (h_{u,d}^{A,B})^0 \rangle$. Their expressions are
\be
    \tan^2 \beta = \frac{\mu^2 + m_{H_d}^2}{\mu^2 + m_{H_u}^2 } ~,  \qquad f^2 = \frac{2}{\lambda^2} \Big( \frac{2b}{\sin 2\beta} - 2\mu^2 - m^2_{H_u} - m^2_{H_d}\Big)\,.
\ee
To capture the leading IR physics, following \cite{Craig:2013fga} we may work in unitary gauge and parameterise the two scalar doublets in the direction of the uneaten pNGB Higgs field as
    \be \label{eq:phiu_phid_IR}
        \phi_u = f \sin\beta \begin{pmatrix}
            \vec{0}_3\\
            s_h \\
            \vec{0}_3\\
            c_h
        \end{pmatrix} ~~,\qquad         \phi_d = f \cos \beta\begin{pmatrix}
            \vec{0}_3 \\
            s_h\\
            \vec{0}_3\\
            c_h
        \end{pmatrix} ~~,
    \ee
where the second Higgs doublet in each sector is assumed to be too heavy to be relevant at the EW scale. We have defined the shorthand notation $s_{h} \equiv \sin\, (h / f)$, $c_{h} = \cos\, (h / f)$.

The novel ingredient we introduce here is a SUSY-breaking spurion coupling which also explicitly breaks the SO(8) global symmetry down to $\text{SO}(4)_A \times \text{SO}(4)_B$. As anticipated, this corresponds to a traceless symmetric irrep we term $F$, which generates the scalar potential
\be
\epsilon\hspace{0.3mm} F^{\,i_1\ldots i_{2n}}\phi_{i_1} \ldots \phi_{i_{2n}} ~~,
\ee
where $\epsilon$ is a small dimensionless parameter and the $\phi_{i_j}$ can be of up-type, down-type or both. Inserting the IR expressions of the fields given in Eq.~\eqref{eq:phiu_phid_IR} we see that the $\beta$-dependence only enters the overall prefactor, while the functional form of the potential is fixed to a Gegenbauer polynomial~\cite{Durieux:2022sgm},
\begin{equation}\label{expbreak}
    V_G^{\left(n\right)}= \epsilon\hspace{0.2mm} H(\beta) \lambda^2 f^4  
    G_n^{3/2}\Big(\hspace{-0.75mm}\cos \frac{2h}{f}  \Big)\;.
\end{equation}
The normalisation is set by the mass-squared $\sim \lambda^2 f^2$ of the $\text{SO}(8)$ radial mode (the Twin Higgs) and $H(\beta)$ is a generic function capturing all the $\beta$ dependence.
 
The EW gauge interactions in the $A$ and $B$ sectors also break $\text{SO}(8)$ explicitly, generating a tree-level $D$-term potential
\begin{equation}
V_D = \frac{g_Z^2 f^4}{32} \cos^2 2\beta\, (s_h^4 + c_h^4)\,,
\end{equation}
where $g_Z \equiv (g^2 + g^{\prime\,2})^{1/2}$. Finally, at one loop the top quark and squarks in each sector generate a Coleman-Weinberg potential, which we evaluate neglecting the $A$-, $\mu$- and $D$-terms. Assuming a single soft mass scale $m^2_{S,A}$ we find for the $A$-sector contribution
\begin{align}\label{eq:V_A}
V_{t + \tilde{t}}^A  \,=\,  \frac{N_c}{64\pi^2} \bigg[ (2m_{S,A}^2 + y_t^2 s_\beta^2 f^2 s_h^2)^2 &\, \Big( \log \frac{2m_{S,A}^2 + y_t^2 s_\beta^2 f^2 s_h^2}{2M^2} - \frac{1}{2} \Big) \nonumber \\
 & - (y_t^2 s_\beta^2 f^2 s_h^2)^2 \Big( \log \frac{y_t^2 s_\beta^2 f^2 s_h^2}{2M^2} - \frac{1}{2} \Big) \bigg]\,,
\end{align}
while the $B$-sector contribution is
\begin{equation}\label{eq:V_B}
V_{t + \tilde{t}}^B = V_{t + \tilde{t}}^A \,(s_h \to c_h, m^2_{S,A}\to m^2_{S,B}) \,.
\end{equation}
Here $M$ is the messenger scale and $s_\beta \equiv \sin \beta$.  Note that $m_t^A = y_t s_\beta v/\sqrt{2}$, hence we take $y_t s_\beta \approx 0.94$ at the top mass scale. The total potential for the pNGB Higgs field is therefore
\begin{equation}
V_h = V_G^{(n)} +  V_D + V_{t + \tilde{t}}^A + V_{t + \tilde{t}}^B\,.
\end{equation}

\subsection*{$\mathcal{Z}_2\,$-$\,$symmetric scenario}
Consider a scenario where the soft masses are $\mathcal{Z}_2$ symmetric, $m_{S,B}^2 = m_{S,A}^2 \equiv m_S^2$. It is informative to expand for $m_S^2 \gg y_t^2 s_\beta^2 f^2 / 2$, exploring the limit where SUSY is truly a UV completion of the low-energy pNGB Higgs dynamics. We find
\begin{equation}\label{eq:Z2_sym}
V_{t + \tilde{t}}^{A+B} (\mathcal{Z}_2\;\mathrm{sym}) + V_D \simeq \frac{N_c (y_t s_\beta)^4 f^4}{64\pi^2} \bigg[s_h^4 \log\frac{a}{s_h^2} + c_h^4 \log\frac{a}{c_h^2} + b_6 (s_h^6 + c_h^6) \bigg],
\end{equation}
where
\begin{equation}\label{eq:matching_EFT}
\log a = \frac{3}{2} + \log \frac{2m_S^2}{y_t^2 s_\beta^2 f^2} + g_Z^2 \cos^2 2\beta \frac{2\pi^2}{N_c y_t^4 s_\beta^4}\,, \qquad b_6 = \frac{y_t^2 s_\beta^2 f^2}{6m_S^2}\,. \\
\end{equation}
We see that, in this limit, even the logarithmic sensitivity to $M$ has canceled. This matching to the Gegenbauer's Twin EFT of~\cite{Durieux:2022sgm} reveals that, even if one minimises the $D$-term contribution by taking $\tan\beta \approx 1$, $\log a$ is too large due to the constant term equal to $3/2$ (recall that in~\cite{Durieux:2022sgm} one needed $\log a < 1$). Furthermore, the $s_h^6 + c_h^6$ term is not negligible. We find that these effects, which follow from retaining $h\,$-$\,$dependent terms in the masses of the UV states (the stops) and the fact that SUSY fixes the value of the interaction strength, prevent a natural electroweak symmetry breaking (EWSB) scale when the Gegenbauer potential $V_G^{(n)}$ is added to Eq.~\eqref{eq:Z2_sym}.

\subsection*{$\mathcal{Z}_2\,$-$\,$breaking scenario}
The inability to generate the desired range of values for $a$ in the $\mathcal{Z}_2$-symmetric case motivates introducing explicit $\mathcal{Z}_2$ breaking in the soft masses, which we parameterise as $\delta m^2 \equiv m_{S,B}^2 - m_{S,A}^2$.  Once added, this breaking allows for viable EWSB.

We now consider the fine-tuning of the model.  The goal is not to argue for a low overall fine-tuning, which is in any case quasi-quantitative in the first place, but instead to adopt one definition in order to understand how the degree of fine-tuning may evolve with an evolving experimental landscape in the future.\footnote{Another reason to avoid an overly quantitative interpretation of the fine-tuning is that the addition of the various ingredients, be they pNGB, SUSY or Gegenbauer, may also be to some extent considered an increase in tuning, since it is an increase in structural elements.} We employ a log-derivative definition of fine-tuning~\cite{Barbieri:1987fn} and define its total size as
\begin{equation}\label{eq:Delta_def}
\Delta = \big[ \mathrm{Tr} (\delta^T \delta) \big]^{-1/2}\,,\qquad \delta = \begin{pmatrix} \frac{\partial \log v^2}{\partial \log \vec{p}} \\ \frac{\partial \log m_h^2}{\partial \log \vec{p}} \end{pmatrix}  ~~,
\end{equation}
where $p_i$ ($i = 1, \ldots, m$) are the chosen input parameters and $\delta$ is a $2\times m$ matrix.  Note that $\delta^T \delta$ is an $m\times m$ matrix, but always has only 2 nonzero eigenvalues.\footnote{Only reparameterisations such that the matrix $T_{ai} = \partial \log q_a / \partial \log p_i$ is orthogonal, where $q_a$ are the new parameters and $p_i$ are the old parameters, leave $\Delta$ invariant, thus there is some parameterisation-dependence in the tuning measure.  This is relatively unimportant for tuning comparisons, but should be kept in mind when considering absolute tuning values.} We emphasise that $\Delta$ only captures the IR component of the tuning; we elaborate further on this point below.

\begin{figure}[t]
\begin{center}
\includegraphics[width=0.49\textwidth]{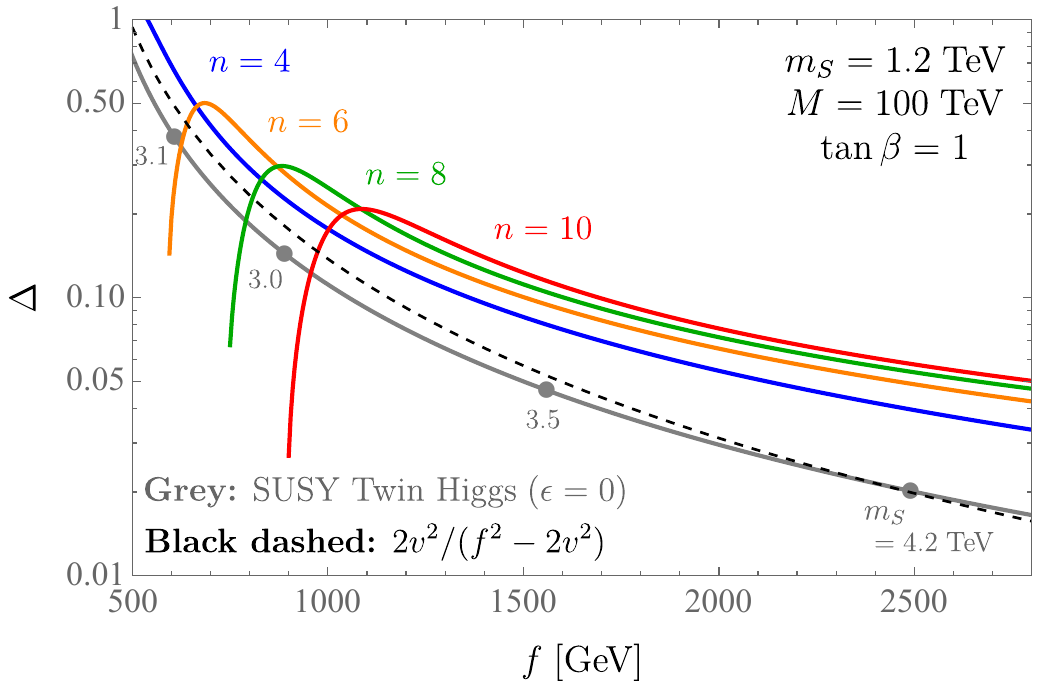}\hspace{1mm}
\includegraphics[width=0.49\textwidth]{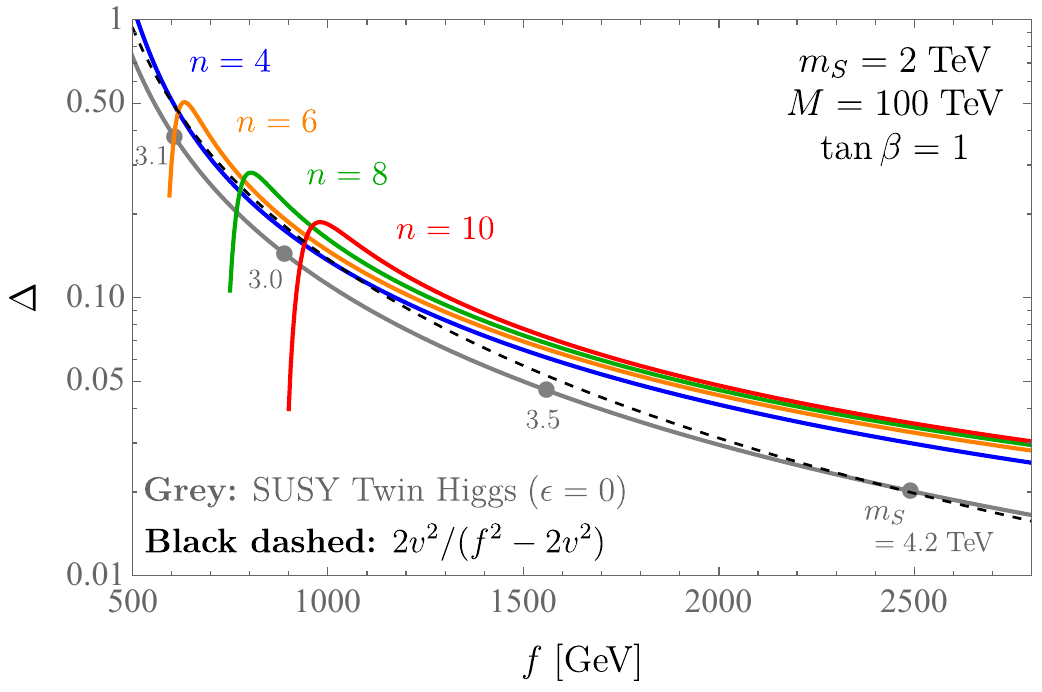}  
\caption{\label{fig:tuning_SUSYGT} Total fine-tuning for the Kitchen Sink model with Gegenbauer index $n = 4, 6, 8$ and $10$, for $M = 100\;\mathrm{TeV}$ and $\tan \beta = 1$. In the left panel we have taken the SM-coloured stop mass scale $m_S = 1.2\;\mathrm{TeV}$, in the right panel $m_S = 2\;\mathrm{TeV}$. The tuning is calculated with respect to $p_i = \{\epsilon, m_S^2, \delta m^2 \}$. The grey line corresponds to the SUSY Twin Higgs model (without Gegenbauer contribution), with tuning calculated with respect to $p_i = \{m_S^2, \delta m^2 \}$. In this model $m_S$ varies for every choice of $f$ and the points indicate representative values. The black dashed line corresponds to $2v^2 / (f^2 - 2 v^2)$.} 
\end{center}
\end{figure}

Figure~\ref{fig:tuning_SUSYGT} shows the absolute IR fine-tuning in the Kitchen Sink model with $\tan\beta = 1$, for a range of decay constants $f$ and two values of the top squark mass scale $m_S$. In both cases it is clear that the model offers some fine-tuning improvement over the na\"ive estimate for a pNGB scenario and also the SUSY Twin Higgs, although an apples-to-apples comparison is lacking in this case as the latter requires different (larger) stop mass values.  Importantly, we see the interplay of the pNGB and SUSY dynamics in keeping the Higgs boson light, with SUSY coming to the rescue for large $f$ and smaller top squark masses. 

\begin{figure}[h!]
\begin{center}
\includegraphics[width=0.49\textwidth]{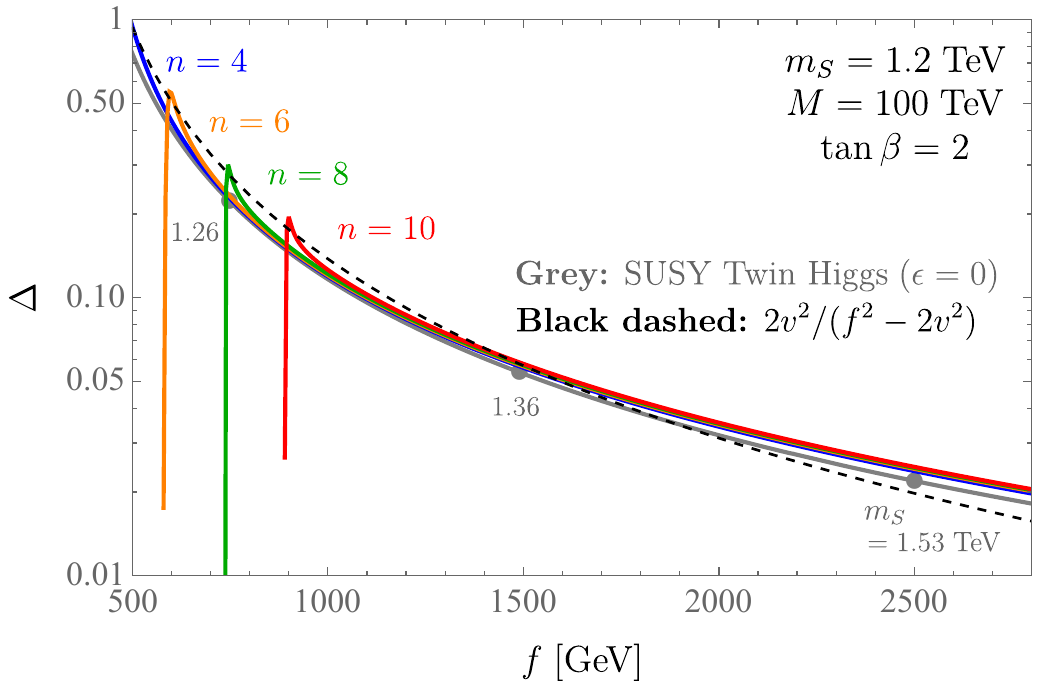}\hspace{1mm}
\includegraphics[width=0.49\textwidth]{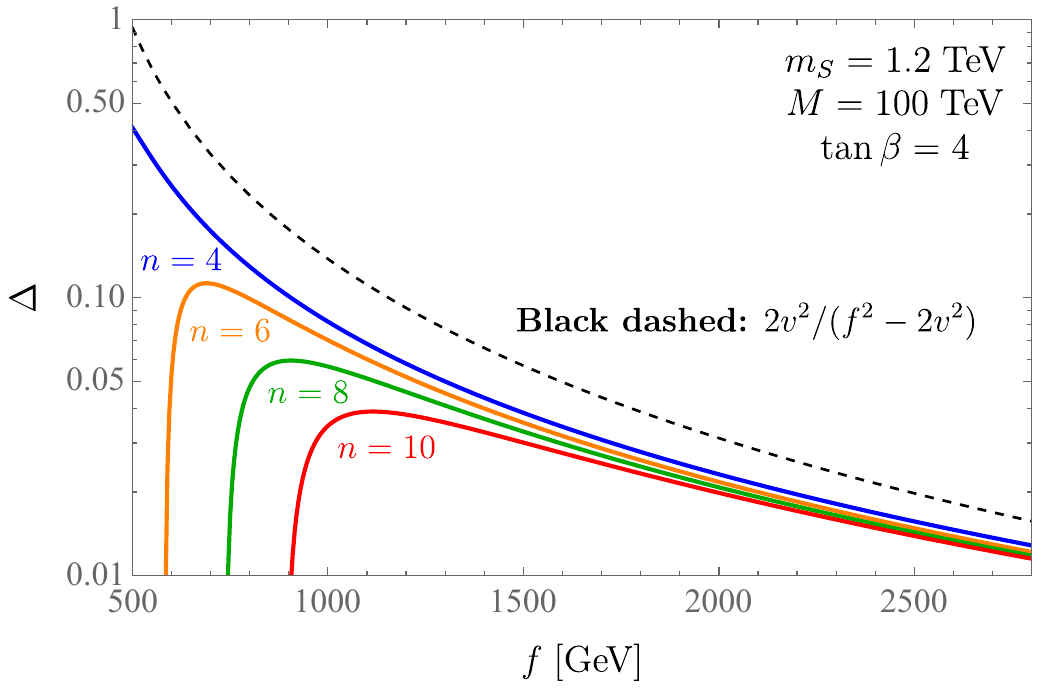}
\caption{\label{fig:tuning_SUSYGT_2} Total fine-tuning for the Kitchen Sink model with Gegenbauer index $n = 4, 6, 8$ and $10$, for $M = 100\;\mathrm{TeV}$ and $m_S = 1.2$~TeV. In the left panel we have taken $\tan\beta = 2$, in the right panel $\tan\beta = 4$. The tuning is calculated with respect to $p_i = \{\epsilon, m_S^2, \delta m^2 \}$. In the left panel, the grey line corresponds to the SUSY Twin Higgs model (without Gegenbauer contribution), with tuning calculated with respect to $p_i = \{m_S^2, \delta m^2 \}$. In this model $m_S$ varies for every choice of $f$ and the points indicate representative values. In the right panel the results for the SUSY Twin Higgs are not shown, because the required SM-coloured stop mass $m_S$ is always well below $1$~TeV and therefore ruled out by LHC searches. The black dashed line corresponds to $2v^2 / (f^2 - 2 v^2)$.} 
\end{center}
\end{figure}

Figure~\ref{fig:tuning_SUSYGT_2} shows that the fine-tuning worsens as $\tan \beta$ is increased, as a result of the larger $D$-term contribution to the Higgs potential. Already for $\tan\beta = 2$ there is no fine-tuning gain relative to a standard SUSY Twin Higgs, whereas for $\tan\beta = 4$ the tuning becomes worse than the baseline expectation $\sim 2v^2 / f^2$. This preference for $\tan\beta \approx 1$ highlights again that in order to be effective in generating naturally $v \ll f$, the Gegenbauer mechanism requires all other contributions to the potential to be as small as possible. We already saw this feature at play in the requirement of small $a$ in the top sector potential, discussed below Eq.~\eqref{eq:a}. Notice that $\tan\beta \approx 1$ requires $m^2_{H_u} \approx m^2_{H_d}$. As already mentioned in~\cite{Craig:2013fga}, this relation between the soft masses could be obtained naturally by adapting the results of~\cite{Falkowski:2006qq}, where a softly broken $\mathcal{Z}_2$ symmetry exchanging $u\leftrightarrow d$ was introduced in the context of a left-right Twin SUSY model. We do not pursue that further model building effort here.


We now return to the issue of possible `UV' tunings that are not captured by our IR definition, by looking more closely at the scales where the low energy EFT we consider is valid and the relationship of the low-energy predictions to the underlying `UV' parameters. Above the scale of the radial modes the 2HDM structure is present in full, hence the input parameters are $b$, $m_{H_u}^2$, $m_{H_d}^2$, $\mu^2$ and $\lambda$. Then $f$, $\tan \beta$ and $m_A^2 = 2b / \sin 2\beta$ are no longer input parameters but predictions, together with $m_h$ and $v$. It may be that fine-tuning is required in order to realise certain values of these parameters.  For instance, obtaining $\tan \beta \approx 1$ may require tuning, unless a natural mechanism is implemented as discussed above. It may also be that some fine-tuning is needed to realise a particular value of $f$ if the stop masses $m_S$ are large \cite{Craig:2013fga}.  For $m_A^2 \gg \lambda^2 f^2$, this tuning scales approximately as
\begin{equation}
\Delta_{f^2} \approx \frac{\lambda^2 f^2 \cos^2 \beta}{\delta m^2_{H_u}} ~~,
\end{equation}
where $\delta m^2_{H_u}$ denotes the radiative corrections to $m^2_{H_u}$ from the stops, which could in principle be large.  Since the Kitchen Sink model naturally accommodates rather large $f$ this tuning is unlikely to be very significant, but it should be kept in mind. Such potential `UV' tunings concern parameters which are, from the perspective of the low energy EFT, inputs. As a result, we do not consider these additional tuning aspects here, since they factor from the tuning of the IR predictions for $m_h$ and $v/f$.  Nonetheless, we note that in the full model containing all radial modes these aspects may impact the overall fine-tuning for a given parameter point.

Finally, we comment on the 2HDM aspects we have neglected in focussing on the low energy EFT. Throughout, we have assumed that the second Higgs doublet of each sector is sufficiently heavy to ensure the alignment limit, so that only the light Higgs doublet needs to be considered at low energies. Increasing the mass of the second Higgs doublet to enforce alignment does not introduce additional tuning in the EW sector: the $\mathcal{Z}_2$ exchange symmetry implies that the quadratic terms of the scalar potential respect the accidental SO(8) invariance, whereas only parameters that break explicitly the $\mathrm{SO}(8)$ can generate a potential for the light pNGB Higgs.

\section{Outlook for Naturalness}\label{sec:Outlook}
\begin{figure}[t!]
\begin{center}
\includegraphics[width=0.666\textwidth]{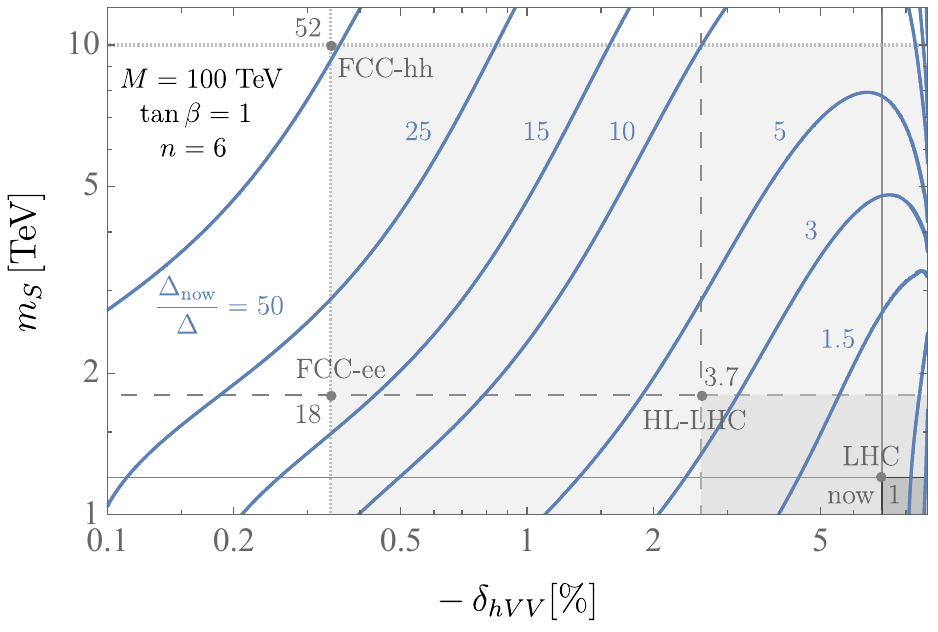}
\caption{Contours of fine-tuning increase, $ \Delta_{\rm now}/\Delta$, for the Kitchen Sink model with Gegenbauer index $n=6$, $M=100$ TeV and $\tan \beta=1$, in the plane of SM-coloured stop mass $m_S$ versus Higgs coupling modification $\delta_{hVV}$. The fine-tuning is calculated with respect to $p_i = \{\epsilon, m_S^2, \delta m^2 \}$. Between now and the FCC-ee~(-hh) era the total range of fine-tuning probed for this model would span a factor $\sim 20$ ($\sim 50$). If the IR definition in Eq.~\eqref{eq:Delta_def} is used to evaluate the present-day tuning, one finds $\Delta_{\rm now} \approx 0.5$.}
\label{fig:contours}
\end{center}
\end{figure}

In Fig.~\ref{fig:contours} we show the projected relative changes in fine-tuning, in the absence of a future discovery, in the plane of SM-coloured stop soft mass $m_S = m_{S,A}$ versus Higgs coupling modification $\delta_{hVV} = (1 - v^2/f^2)^{1/2} - 1$ for the Kitchen Sink model. In the lower-right corner the parameter range that is (approximately) excluded at present is shown, alongside projections for the HL-LHC phase, followed by FCC-ee and FCC-hh. The $v^2$ (or, equivalently, $v/f$) tuning is more significant than the $m_h^2$ tuning, but both contribute appreciably to the total measure $\Delta$ across this parameter space.

Between now and the end of the HL-LHC era, in the absence of any hints for new physics, the fine-tuning would increase by a factor $\sim 4$, driven almost entirely by the improvement of the exclusion from Higgs coupling measurements to $|\delta_{hVV}| < 2.6\%$ ($2\hspace{0.2mm}\sigma$,~\cite{deBlas:2019rxi}). This demonstrates that, for this model, it is Higgs coupling measurements which will be a driving factor for the question of naturalness in the mid-term.  As argued in Sec.~\ref{sec:intro}, this reflects general expectations for other models.

On the other hand, despite an expected improvement in Higgs coupling precision by an order of magnitude, FCC-ee would probe the naturalness question only a factor $5$ more deeply. The reason for this is interesting and exposes an important aspect of naturalness in the time of a precision programme. On the one hand, a Higgs coupling modification smaller than the expected FCC-ee exclusion of $0.34\%$ ($2\hspace{0.2mm}\sigma$,~\cite{deBlas:2019rxi}) corresponds to a compositeness scale $f  > 3.0$ TeV.  On the other hand, the direct stop reach would not extend beyond that of the HL-LHC, $m_S \gtrsim 1.8$~TeV. As a result, in this parameter range the Kitchen Sink model can revert to an almost supersymmetric setup, addressing naturalness through SUSY rather than a pNGB solution, the light stops coming to the rescue.

Extrapolating deeper into the future, FCC-hh would further constrain $m_S \gtrsim 10$ TeV~\cite{EuropeanStrategyforParticlePhysicsPreparatoryGroup:2019qin}, putting paid to any hope for naturalness, Kitchen Sink included, with the tuning estimate exceeding a factor $50$ relative to the present day despite the all-out model-building gymnastics. Ultimately, even for the Kitchen Sink model which, by design, attempts to implement naturalness at any model-building cost, the fine-tuning would be forced at least to the sub-percent level by the combination of FCC-ee and FCC-hh.

\section{Summary}
A natural theory for the Higgs would be one in which its vev and its mass are radiatively stable and calculable within a putative UV completion. In practice, this translates into a present-day tension that arises from both the absence of new resonances and the lack of Higgs coupling deviations. This tension will only increase if no new physics is uncovered by the end of the HL-LHC phase. Rough quantitative estimates indicate that, in pNGB Higgs scenarios, the relative reach of direct versus indirect probes will increasingly tilt towards the latter as driving the tension.

By building a `Kitchen Sink' model, we have attempted to characterise and explore the broad landscape of symmetry-based naturalness solutions. The key ingredients may be summarised as follows. In pNGB Higgs scenarios it is usually assumed that the only sources of explicit symmetry breaking are those coming from the SM Yukawa and gauge interactions. Going beyond minimality, we may introduce an additional source of explicit breaking in the form of a spurion in a higher order irreducible representation of the global symmetry group. This generates a Gegenbauer potential~\cite{Durieux:2021riy} which can shift the global minimum to a small field value, realising more naturally a $v/f \ll 1$ hierarchy. From the perspective of Higgs coupling modifications, which scale as $v^2/f^2$, this would mean that a Gegenbauer pNGB Higgs is naturally more SM-like. However, even though the Gegenbauer Higgs mechanism somewhat ameliorates the tuning as compared to standard pNGB models, it does not address the absence of new SM-coloured resonances. Twinning the Gegenbauer \cite{Durieux:2022sgm} helps with this second aspect, but comes with an extra requirement: $\log a < 1$, where the parameter $a$ is related to the coupling of new coloured resonances in the top sector as $a \approx g_\ast^2 / y_t^2$. This points towards a perturbative theory, such as SUSY, as a possible UV completion. In considering the SUSY Twin setup, in contrast with previous treatments in the literature~\cite{Craig:2013fga}, our analysis fully retains the Higgs dependence in the one-loop effective potential. We find that, as a result, $\log a$ in the Kitchen Sink model remains too large, requiring the introduction of an extra source of Twin symmetry breaking in the stop soft masses in order to achieve viable EWSB.

We estimated the total IR fine-tuning for the Kitchen Sink model with respect to the relevant high-scale parameters. The Gegenbauer contribution ameliorates the tuning relative to a standard SUSY Twin Higgs model, particularly for large values of $f$ (see Fig.~\ref{fig:tuning_SUSYGT}), and it also improves upon the $v/f$ tuning expectation for more generic composite Higgs models. At present, the IR tuning can still be of $\mathcal{O}(1)$: Fig.~\ref{fig:contours} shows that $\Delta \approx 0.5$ is still possible, though only just. However, we have found that $\tan\beta \approx 1$ is a necessary condition for the fine-tuning gain: the tree-level \mbox{$D$-term} potential must be suppressed, as the Gegenbauer mechanism is effective only when all other contributions to the pNGB Higgs potential are sufficiently small.

Fig.~\ref{fig:contours} summarises the main focus of this paper, namely the broad phenomenological prospects for the naturalness question between now and the end of a putative FCC-hh era. From our discussion of the Kitchen Sink model we conclude that in the mid-term future, in the absence of signals for new physics, the fine-tuning would be dominated by precision measurements probing the $\mathcal{O}(v^2/f^2)$ contribution to Higgs couplings. Ultimately, the fine-tuning would be exacerbated by a factor $\sim 50$, down to at least the $1\%$ level, despite the numerous field theory tools simultaneously thrown at naturalness in this model. 

In summary, we argue that for pNGB approaches to Higgs naturalness we are presently astride the `direct search' and `indirect precision' eras, with the latter experimental strategy set to dominate progress on the naturalness frontier in the near- and mid-term. If a future Higgs factory such as FCC-ee is realised then precision Higgs coupling measurements would probe significantly deeper. More optimistically, it would seem that if evidence of naturalness is to arise in the foreseeable future then it may emerge first in the form of Higgs coupling deviations.

\label{sec:sum}

\acknowledgments{AM acknowledges support by the DFG Emmy Noether Grant No.~PI 1933/1-1, by the DFG under Germany’s Excellence Strategy - EXC 2121 Quantum Universe - 390833306, and by the National Research Foundation of Korea~(NRF) grant funded by the Korea government~(MSIT)~(No.~2012K1A3A2A0105178151). ES was supported by the Science and Technology Facilities Council under the Ernest Rutherford Fellowship ST/X003612/1.
}



\bibliographystyle{apsrev4-1_title}
\bibliography{biblio}

\end{document}